\documentclass[preprint,amsmath,amssymb,aip,superscriptaddress,nofootinbib,showkeys]{revtex4-2}

\usepackage[english]{babel}
\usepackage[utf8]{inputenc}
\usepackage{xr-hyper}

\usepackage[colorlinks]{hyperref}
\usepackage[T1]{fontenc}
\usepackage[dvips]{graphicx}
\usepackage{amsmath}
\usepackage{amsfonts}
\usepackage{color}
\usepackage[caption=false, justification=justified]{subfig}
\usepackage{url}
\usepackage{ulem}

\makeatletter
\newcommand*{\addFileDependency}[1]{
  \typeout{(#1)}
  \@addtofilelist{#1}
  \IfFileExists{#1}{}{\typeout{No file #1.}}
}
\makeatother

\newcommand*{\myexternaldocument}[1]{%
    \externaldocument{#1}%
    \addFileDependency{#1.tex}%
    \addFileDependency{#1.aux}%
}

\myexternaldocument{suppmat_REV2_1502}

\hypersetup{filecolor=blue}
\hypersetup{citecolor=blue}
\hypersetup{urlcolor=blue}
\hypersetup{linkcolor=blue}

\begin{document}

\title{Spintronic THz emitters based on transition metals and semi-metals/Pt multilayers.}
\date{\today}

\author{J.~Hawecker}
\affiliation{Laboratoire de Physique de l’Ecole Normale Supérieure, ENS, Université PSL, CNRS, Sorbonne Université, Université de Paris, F-75005 Paris, France}
\author{E.~Rongione}
\affiliation{Unité Mixte de Physique, CNRS, Thales, Université Paris-Saclay, F-91767 Palaiseau, France}
\affiliation{Laboratoire de Physique de l’Ecole Normale Supérieure, ENS, Université PSL, CNRS, Sorbonne Université, Université de Paris, F-75005 Paris, France}
\author{A.~Markou}
\affiliation{Max Planck Institute for Chemical Physics of Solids, D-01187 Dresden, Germany}
\author{S.~Krishnia}
\affiliation{Unité Mixte de Physique, CNRS, Thales, Université Paris-Saclay, F-91767 Palaiseau, France}
\author{F.~Godel}
\affiliation{Unité Mixte de Physique, CNRS, Thales, Université Paris-Saclay, F-91767 Palaiseau, France}
\author{S.~Collin}
\affiliation{Unité Mixte de Physique, CNRS, Thales, Université Paris-Saclay, F-91767 Palaiseau, France}
\author{R.~Lebrun}
\affiliation{Unité Mixte de Physique, CNRS, Thales, Université Paris-Saclay, F-91767 Palaiseau, France}
\author{J.~Tignon}
\affiliation{Laboratoire de Physique de l’Ecole Normale Supérieure, ENS, Université PSL, CNRS, Sorbonne Université, Université de Paris, F-75005 Paris, France}
\author{J.~Mangeney}
\affiliation{Laboratoire de Physique de l’Ecole Normale Supérieure, ENS, Université PSL, CNRS, Sorbonne Université, Université de Paris, F-75005 Paris, France}
\author{T. Boulier}
\affiliation{Laboratoire de Physique de l’Ecole Normale Supérieure, ENS, Université PSL, CNRS, Sorbonne Université, Université de Paris, F-75005 Paris, France}
\author{J.-M.~George}
\affiliation{Unité Mixte de Physique, CNRS, Thales, Université Paris-Saclay, F-91767 Palaiseau, France}
\author{C.~Felser}
\affiliation{Max Planck Institute for Chemical Physics of Solids, D-01187 Dresden, Germany}
\author{H.~Jaffrès}
\affiliation{Unité Mixte de Physique, CNRS, Thales, Université Paris-Saclay, F-91767 Palaiseau, France}
\affiliation{\normalfont Corresponding authors:~\href{mailto:henri.jaffres@cnrs-thales.fr}{henri.jaffres@cnrs-thales.fr},~\href{mailto:sukhdeep.dhillon@phys.ens.fr}{sukhdeep.dhillon@phys.ens.fr}}
\author{S.~Dhillon}
\affiliation{Laboratoire de Physique de l’Ecole Normale Supérieure, ENS, Université PSL, CNRS, Sorbonne Université, Université de Paris, F-75005 Paris, France}
\affiliation{\normalfont Corresponding authors:~\href{mailto:henri.jaffres@cnrs-thales.fr}{henri.jaffres@cnrs-thales.fr},~\href{mailto:sukhdeep.dhillon@phys.ens.fr}{sukhdeep.dhillon@phys.ens.fr}}

\newpage

\begin{abstract}
Spintronic terahertz (THz) emitters (STE) based on the inverse spin Hall effect in ferromagnetic/heavy metal (FM/HM) heterostructures have become important sources for THz pulse generation. The design, materials and control of these interfaces at the nanometer level has become vital to engineer their THz emission properties. In this work, we present studies of the optimization of such structures through a multi-pronged approach, taking advantage of material and interface engineering to enhance the THz spintronic emission. This includes: the application of multi-stacks of HM/FM junctions and their application to trilayer structures; the use of spin-sinks to simultaneously enhance the THz emitted fields and reduce the use of thick Pt layers to reduce optical absorption; and the use of semi-metals to increase the spin polarization and thus the THz emission. Through these approaches, significant enhancements of the THz field can be achieved. Importantly, taking into account the optical absorption permits to elucidate novel phenomena such as the relation between the spin diffusion length and the spin-sink using THz spectroscopy, as well as possibly distinguishing between self– and interface- spin-to-charge conversion in semi-metals.
\end{abstract}


\maketitle


Spintronic terahertz (THz) emitters (STE) have considerably impacted THz technology in a remarkably short time frame~\cite{Seifert2016,bull2021,cheng2021,Papaioannou2020}, becoming a viable and alternative source to nonlinear crystals and photoconductive antennas for THz time-domain spectroscopy (TDS). These structures typically consist of nanometer thick ferromagnetic (FM) - heavy metal (HM) junctions. When optically excited with a femtosecond near-infrared (NIR) laser, an out-of-equilibrium spin current is generated in the former, the so-called superdiffusive flux~\cite{battiato2010,battiato2012}, that is converted to a transient charge-current in the latter - spin-to-charge conversion (SCC) - \textit{via} the inverse spin Hall effect (ISHE). This current then gives rise to a linearly polarized THz pulse where the polarization can be easily controlled by a small applied magnetic field. Compared to other established technologies, THz spintronic sources have the advantages of generating phonon-less THz spectra, being extremely thin, with performances surpassing those of thick (hundreds of microns) nonlinear crystals. As well as an important THz technology, THz emission TDS has now proven to be an essential tool to probe ultrafast spintronic phenomena at FM/HM interfaces as well as in a range of novel materials, from topological insulators to transition metal dichalcogenides~\cite{Cheng2019,Yagodkin2021}.

While FM/HM based interfaces such as W/Co$_{40}$Fe$_{40}$B$_{20}$/Pt~\cite{Seifert2016} represent the materials of reference, this letter presents routes towards THz emitters using engineered spintronic interfaces to reach higher THz fields, as well as enhancing our understanding of THz SCC. More specifically, beyond tuning the excitation wavelength~\cite{Adam2019,Herapath2019,Papaioannou2018,Beigang2019}, improvements in the THz emission efficiency \textit{via} the ISHE in HM based junctions are anticipated by considering three major strategies: \textit{i}) metallic multi-stacks of spintronic junctions of high electronic transparency~\cite{Dang2020,yuan2020,Hawecker2021,wahada2021} allowing a large number of emitting dipole planes, but sufficiently thin to minimise NIR and THz optical absorption; \textit{ii}) by optimized stacks coupled with specific \textit{spin-sinks} such as Au:W alloys~\cite{lackzowski2014,lackzowski2016,lackzowski2017,dangjaffres2020} to avoid detrimental hot spin reflection, where we show theoretically and experimentally in the limit of thin films ($\approx$ 2~nm) a large gain in the THz E-field; \textit{iii}) by means of optically excited spin-polarization enhancement with the use of semi-metals such as those based on Heusler compounds (presently Co$_2$MnGa) possessing an intrinsic high spin-polarization.

\begin{figure*}[!ht]
\begin{center}
\includegraphics[width=\textwidth]{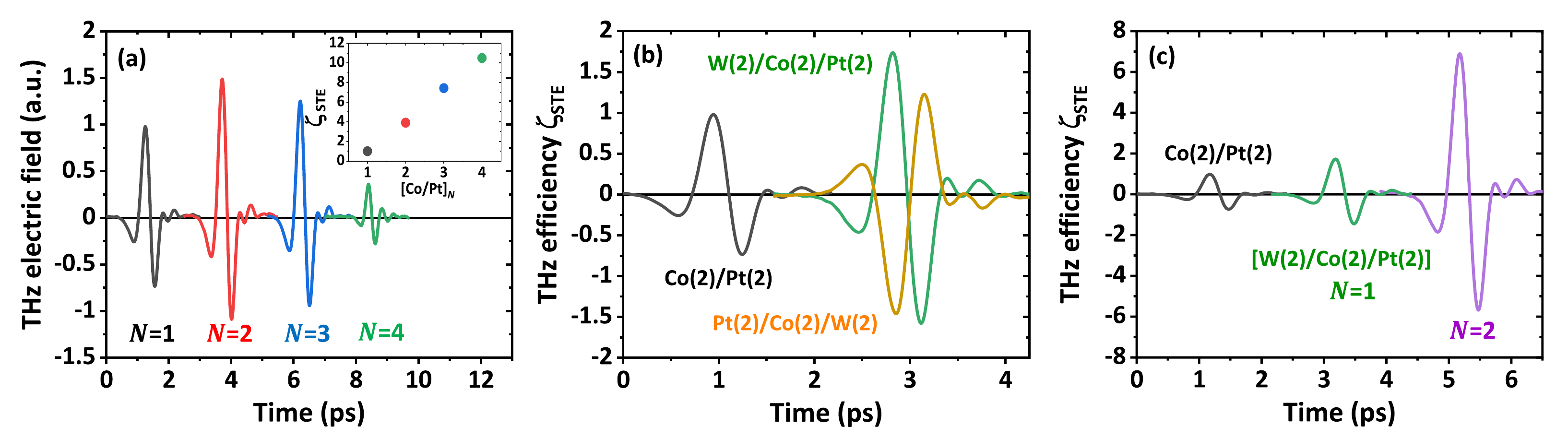}
\caption{\textbf{Periodic stack of STE metallic layers.} (a) Raw THz emission from glass//[Co(2)/Pt(2)]$_N$ multistacked layers separated with AlO$_x$(1). Time traces are shifted for clarity. Inset: Withdrawal of NIR and THz contribution leading to the STE electronic efficiency $\zeta_\text{STE}$ \textit{vs.} the $N$-th Co/Pt repetition. (b) Example of $\zeta_\text{STE}$ for in-phase dual SCC with W/Co/Pt trilayers where the negative spin Hall angle of W participates to THz emission enhancement. (c) $\zeta_\text{STE}$ obtained by taking into account the THz and NIR absorption for Co(2)/Pt(2), [W(2)/Co(2)/Pt(2)]$_1$ and [W(2)/Co(2)/Pt(2)]$_2$ separated by AlO$_x$(1).}
\label{fig1}
\end{center}
\end{figure*}

In this paper, by exploiting the properties of transition metal interfaces, we explore these three different aspects and show that the STE electronic efficiency $\zeta_\text{STE}$ may be enhanced by at least a factor $\simeq$2 for the emitted THz field and thus power enhancements by a factor $\simeq$4. Importantly, to explain these enhancements, we analyse our results when stripped from the effect of optical (NIR and THz) absorption to highlight the intrinsic physics of SCC at play. In particular, the THz absorption plays a detrimental role in the signal reduction by a typical factor lying between 2 and 4 (output power reduced by one order of magnitude). In all experiments, the STEs are excited by a Ti:Sa oscillator delivering $\simeq$100 fs pulses centered at 810 nm and the emitted THz pulses are detected using electro-optic sampling in a THz-TDS system (see~Suppl.~Mat.~\textcolor{blue}{S1}).

\vspace{0.1in}

As a proof of concept, our first approach to investigate THz emission of STE FM/HM structures is to grow periodic stacks of thin Co/Pt layers to increase the total emitting dipoles and hence the ultrafast spin-current generation~\cite{yang2016,feng2018,Schneider2022}. [Co(2)/Pt(2)]$_N$ multi-stack samples were grown on glass substrates with the number in parenthesis indicating the thickness of the layer in nanometers and $N$ the number of periods (1, 2, 3 and 4) with a 1~nm thick AlO$_x$ separating each period in order to avoid opposite charge dipoles within the structures (details in Suppl.~Mat~\textcolor{blue}{S1}). The emitted THz fields $E_\text{THz}$ are shown in Fig.~\ref{fig1}\textcolor{blue}{a} showing that $N=2$ provides the strongest $E_\text{THz}$ by a factor 1.5 in respect to the $N=1$ reference. The THz signal decreases for $N=3, 4$ by a factor $\simeq$1.25, 5 respectively which can be understood by a larger NIR and THz absorption on increasing the periods. This indicates that $N=2$ corresponds to the best compromise between the added dipoles planes contributing to $E_\text{THz}$ and the wave absorption. The withdrawal of NIR and THz absorption contributions to the THz emission for each repetition period (Suppl.~Mat.~\textcolor{blue}{S3} and associated references~\cite{meinert2020,gueckstock2021,Rongione2022}) as shown in the inset of Fig.~\ref{fig1}\textcolor{blue}{a} allows the introduction of the \textit{STE electronic efficiency} $\zeta_\text{STE}$. The latter quantity refers, also for the forward discussions, to the effective spin relaxation length in Pt (HM) for the hot electron spin-current. Interestingly, this shows a linear behaviour of the THz field for increasing number of Co/Pt period and indicates that the limiting factor is the optical absorption for the THz generation.

As shown by Seifert \textit{et.~al.}~\cite{Seifert2016}, another approach towards signal enhancement is to design trilayers involving a FM layer embedded between two HMs of opposite spin Hall angle signs (\textit{resp.} Pt and W), thus harvesting opposite spin currents. Here the different contributions of SCC from each HM interface interfere constructively, resulting in an enhanced $E_\text{THz}$. Fig.~\ref{fig1}\textcolor{blue}{b} compares the THz pulses compensated from NIR and THz absorption from W(2)/Co(2)/Pt(2) and Pt(2)/Co(2)/W(2) samples grown by the same method to a Co(2)/Pt(2) reference. The THz signal reaches factors of $\simeq$1.7 and $\simeq$1.4 for W(2)/Co(2)/Pt(2) and Pt(2)/Co(2)/W(2) multilayers respectively. As expected by the spin Hall angle of the first metal excited by the fs oscillator (Pt in W(2)/Co(2)/Pt(2) and W in Pt(2)/Co(2)/W(2)), the phase of the THz pulse reverts accordingly. Merging now the periodic stacks and trilayers concepts, samples of [W(2)/Co(2)/Pt(2)]$_2$ with 1 nm AlO$_x$ separators between the periods were realized. The results shown in Fig.~\ref{fig1}\textcolor{blue}{c} clearly show a STE electronic efficiency enhanced by a factor $\simeq$7.5 with respect to the Co/Pt reference when the optical absorption is withdrawn. This layer repetition approach demonstrates the crucial limitation of both NIR and THz absorption in STE structures. The same overall qualitative and quantitative conclusions are raised by replacing Co by Co$_{40}$Fe$_{40}$B$_{20}$ by comparing results from [W/Co$_{40}$Fe$_{40}$B$_{20}$/Pt]$_1$ and [W/Co$_{40}$Fe$_{40}$B$_{20}$/Pt]$_2$ trilayers as shown on Fig.~\ref{fig1}\textcolor{blue}{c}.

The modeling of the THz emission of periodic structures requires the expression of $E_\text{THz}$ emitted from the multiple dipole planes according to Refs.~\cite{gueckstock2021,meinert2020} and Suppl.~Mat.~S2 which includes multiple reflections consideration in the multilayered structures:

\begin{eqnarray}
    E_{\text{THz}}(N)\simeq \frac{Z_0 ~N_m \left[1-\exp\left(-N/N_m\right)\right]}{1+n_\text{sub}+N ~Z_0 \int \sigma_p dz_p} \theta_{\text{SHE}} ~\zeta_{\text{STE}}
    \label{multiple}
\end{eqnarray}
where $Z_0= 377~\Omega$ is the free space impedance, $n_\text{sub}$ is the optical index of the substrate, $\int \sigma_p dz_p$ is the conductivity of a single period integrated over its thickness and $N_m=\lambda_{opt}/d_p$ the maximum numbers of the dipole planes excited by optical absorption (ratio of the NIR absorption depth $\lambda_{opt}\approx 12$~nm~\cite{palik1998} over the metallic multilayer period $d_p$, presently 6~nm). The optimum $N=\tilde{N}$ is obtained by minimizing Eq.~\ref{multiple} over $N$ giving \textit{in fine} $\frac{\tilde{N}}{N_m}=\ln \left(1+\frac{1+n_\text{sub}}{Z_0\int \sigma_p dz_p}+\frac{\tilde{N}}{N_m}\right)\simeq$1 once considered the numerical parameters. The calculation gives an optimal value of $\tilde{N}\simeq 2.3$ in agreement with experiments ($\tilde{N}\simeq 2$).

 \vspace{0.1in}

\begin{figure*}[!ht]
\begin{center}
\includegraphics[width=1\textwidth]{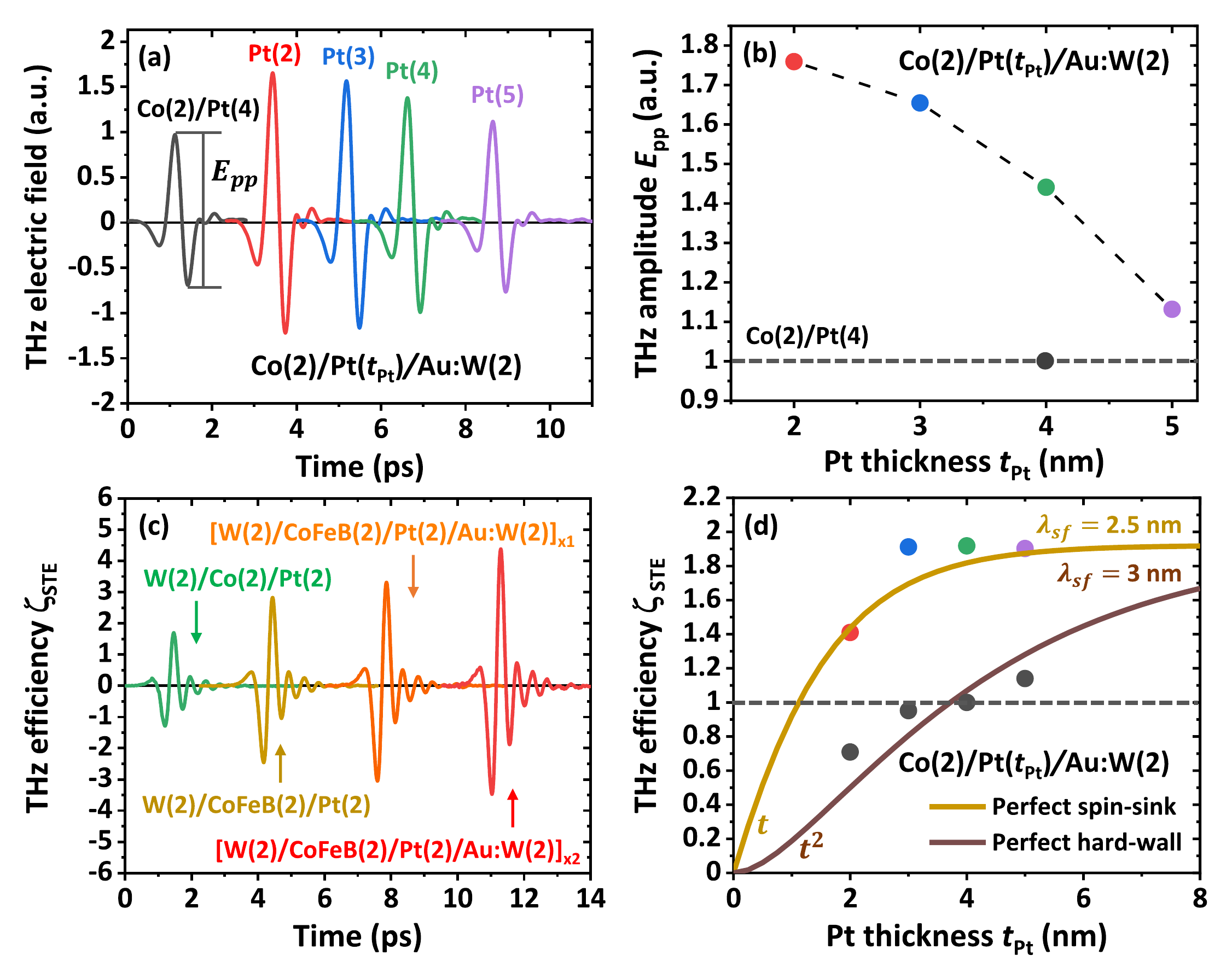}
\caption{\textbf{Au:W spin-sink for STE.} THz emission of Co(2)/Pt($t_{{\text{Pt}}}$)/Au:W(2) \textit{vs.}~Pt thickness $t_{{\text{Pt}}}$ (resp.~2,~3,~4,~5~nm) represented with (a) THz pulse generation in the time-domain (Time traces are shifted in time for clarity.) and (b) the peak amplitude, normalized with respect to the emission from Co(2)/Pt(4). The thin dashed line is a guide to the eyes. (c) THz efficiency $\zeta_\text{STE}$ presented in the case of trilayer emitter of W(2)/Co(2)/Pt(2) (green) compared to \textit{i}) Co$_{40}$Fe$_{40}$B$_{20}$ as the spin-injection material (red) and \textit{ii}) \textit{spin-sink} trilayer emitter with $N=1,2$ repetition(s) (respectively in orange and red). All the efficiencies are presented normalized with respect to Co(2)/Pt(2). (d) Renormalized peak amplitude taking into consideration NIR and THz absorption, illustrating the STE electronic efficiency $\zeta_{\text{STE}}$ \textit{vs.} Pt thickness $t_\text{Pt}$. Perfect \textit{spin-sink} (shown in yellow) and \textit{hard-wall} (shown in brown) limits are derived from Eq.~\eqref{zeta}.}
\label{fig2}
\end{center}
\end{figure*}

The second part of this work is the application of metallic \textit{spin-sinks}~\cite{Legrand2016} as a method to optimize STEs. Here, we propose the use of Au:W alloys possessing the advantage of a high spin interface transparency \cite{Hawecker2021,gupta2020} with Pt along with a large resistivity ($\rho_{\text{AuW}} = 87-130~\mu\Omega$.cm). This makes Au:W alloys perfect candidates to avoid spin backflow at the outward Pt interface. More details about the Au:W allow deposition can be found in Suppl.~Mat.~\textcolor{blue}{S1}. The samples studied were a Co(2)/Pt($t_{\text{Pt}}$)/Au:W(2) series deposited on MgO($001$). Fig.~\ref{fig2}\textcolor{blue}{a} displays the acquired THz pulses mapped by THz-TDS for different Pt thickness $t_{{\text{Pt}}}=2, 3, 4$ and 5~nm; Fig.~\ref{fig2}\textcolor{blue}{b} displays the THz peak-to-peak amplitude $E_{pp}$ for each sample. THz pulses are compared to the reference Co(2)/Pt(4) grown within the same batch. The THz emission from Co(2)/Pt(4)/Au:W(2) is almost 50\% larger than the emission from Co(2)/Pt(4) one, despite the latter being characterized by having a smaller optical absorption (both THz and NIR), illustrating the significant \textit{spin-sink} effect (SSE) using Au:W in the sub-picosecond domain. A subsequent increase of the THz emission efficiency is also observed for the whole series $t_{\text{Pt}}=2, 3, 5$~nm (Fig.~\ref{fig2}\textcolor{blue}{d}). Moreover, by comparing the results obtained for different Pt thicknesses, one evidences an even larger emission efficiency for $t_{{\text{Pt}}}=2$~nm by almost 80\% compared to Co(2)/Pt(4). These results are in favor of a strong SSE enhanced by the ability of reducing the THz absorption through the reduction of the Pt thickness of a large conductance ($\rho_{{\text{Pt}}} \approx 30~\mu \Omega$.cm). The drop of the THz pulse amplitude for $t_{{\text{Pt}}}=3,~4,~5$~nm from its maximal value follows from a smaller SSE as $t_\text{Pt}$ exceeds its spin-relaxation length as well as to a larger THz absorption (THz and NIR pump). Fig.~\ref{fig2}\textcolor{blue}{c} presents the renormalized THz emission after having taken into account the THz and NIR absorptions in the layered structures. This gives the STE electronic efficiency $\zeta_\text{STE}$. We notice that below the electronic spin relaxation length (3~nm for Pt), the conversion builds-up until it saturates after the relaxation length: this expected behaviour is in line with ISHE. At comparable structure thickness (\textit{i.e.}~Pt(4)), the addition of Au:W(2) allows to increase the conversion efficiency by nearly a factor of 2 in field. A particular interesting point is the observation of a same increase of $\zeta_{\text{STE}}$ with the inclusion of Au:W spin-sink in [W(2)/CoFeB(2)/Pt(2)/Au:W(2)]$_{2}$ compared to W(2)/CoFeB(2)/Pt(2)/Au:W(2) owing to avoiding backward majoritary spin-current reflection.

\vspace{0.1in}

The effect of the \textit{spin-sink} layer can be modelled with the assumption of the existence of effective momentum and spin relaxation lengths in the THz domain in the wave-diffusion approach~\cite{kaltenborn2012,nenno2019,Dang2020}. More details on the ultrafast electron diffusion model are discussed in Suppl.~Mat.~\textcolor{blue}{S3}. We are searching the STE efficiency $\zeta_{{\text{STE}}}$ and subsequent SCC integrated over the Pt thickness. The expression of $\zeta_{{\text{STE}}}$ can be generalized in terms of a three-layers model and may found for the two limiting cases of \textit{i}) a perfect \textit{hard-wall} or perfect (detrimental) spin reflections at the top Pt and \textit{ii}) a perfect \textit{spin-sink} as approached by Au:W yielding \textit{in fine} to:

\begin{widetext}
\begin{equation}
\zeta_{\text{STE}}^{\text{ss}} \underset{\text{spin-sink}}{=} \frac{ ~\lambda_{sf}^{\text{Pt}}}{\coth\left(\frac{t_{\text{Pt}}}{\lambda_{sf}^{\text{Pt}}}\right)+\frac{r^s_\text{Pt}}{r^s_\text{F}}} \qquad \text{and} \qquad \zeta_{\text{STE}}^{\text{hw}} \underset{\text{hard-wall}}{=} \frac{\lambda_{sf}^{\text{Pt}}}{\coth\left(\frac{t_{\text{Pt}}}{2\lambda_{sf}^{\text{Pt}}}\right)+\frac{r^s_\text{Pt}}{2 r^s_\text{F}} \left[1+\coth^2\left(\frac{t_{\text{Pt}}}{2\lambda_{sf}^{\text{Pt}}}\right)\right]}
    \label{zeta}
\end{equation}
\end{widetext}
with $\lambda_{sf}^{\text{Pt}}$ the spin-relaxation length, the volume of the time-oscillating dipoles.

The perfect \textit{spin-sink} (ss) corresponds to a zero spin-resistance~\textit{i.e}~$r_{{\text{spin-sink}}}^s\approx0$ that is no spin-accumulation at its interface with Pt. The perfect \textit{hard-wall} (hw) limit corresponds to an infinite spin-resistance $r_{\text{hard-wall}}^s \approx \infty$ giving zero spin-current and a maximum spin-accumulation. The two expressions of Eq.~\eqref{zeta} match well the variation of the renormalized THz emission experimentally observed in Fig.~\ref{fig2}\textcolor{blue}{c} with $\lambda_{sf}^{{\text{Pt}}}\approx 2.75\pm 0.25$~nm, with $r_{\text{F}}^s\simeq 3$ and $r_{{\text{Pt}}}^s\simeq 2$ the respective Co and Pt spin-resistance (Ref.~\cite{jaffres2020}) in unit of the inverse of the Sharvin resistance ${G_{{\text{sh}}}^{-1}}$. The positive action of the \textit{spin-sink} is obtained by considering the expression of $\zeta_\text{STE}$ (Eq.~\eqref{zeta}) and its dependence \textit{vs.} $t_{\text{Pt}}$. The quantitative gain in $\zeta_{\text{STE}}$ can be also traced by considering the figure of merit $\mathcal{G}^\infty=\left(\zeta_{\text{STE}}^\text{ss}-\zeta_{\text{STE}}^\text{hw}\right)/\zeta_{\text{STE}}^\text{hw}$:
\begin{equation}
\mathcal{G}^\infty=\frac{r_{\text{Pt}}^s \coth\left(\frac{t_{\text{Pt}}}{2\lambda_{sf}^{\text{Pt}}}\right)+r_\text{F}^s}{r_\text{F}^s\cosh\left(\frac{t_{\text{Pt}}}{\lambda_{sf}^{\text{\text{Pt}}}}\right)+r_{\text{Pt}}^s \sinh\left(\frac{t_{\text{Pt}}}{\lambda_{sf}^{\text{Pt}}}\right)}>0
\end{equation}
which manifests a gain ratio in the spin-injection efficiency $\frac{\zeta_{STE}^{ss}}{\zeta_{STE}^{hw}}=1+\mathcal{G}^\infty>1$.

The gain in the STE electronic efficiency goes to zero in the limit of infinite Pt thickness, $t_{\text{Pt}}\gg \lambda_{sf}^{\text{Pt}}$. On the other hand, $\mathcal{G}^\infty$ goes to infinite in the limit of small $t_{\text{Pt}}$. This is due to a linear increase of $\zeta_{\text{STE}}$ in $t_{\text{Pt}}$ for a perfect \textit{spin-sink} whereas only a quadratic increase of $\zeta_{\text{STE}}$ in $t_{\text{Pt}}$ is expected for a \textit{hard-wall} system. As an example, we have plotted on Fig.~\ref{fig2}\textcolor{blue}{c} the expected $\zeta_{\text{STE}}$ obtained from the two opposite situations of a perfect \textit{spin-sink} and of a \textit{hard-wall} for $t_{\text{Pt}}$ at the vicinity of $\lambda_{sf}^{\text{Pt}}$. Fig.~\ref{fig2}\textcolor{blue}{c} and the corresponding guidelines express that Co(2)/Pt(1)/Au:W(2) gives an equivalent signal in size to Co(2)/Pt(4) from a pure electronic efficiency point-of-view with, however, the benefit of a reduced THz absorption. The results is also true for Co(2)/Pt(2)/Au:W(2). It leads to the extraction of a spin diffusion length $\lambda_{sf}^\text{Pt} = 2.75 \pm 0.25$ nm on MgO in close agreement with values assigned in the literature for the present Pt conductivity~\cite{jaffres2014,casanova2016,berger2018}. The field amplitude $E_{\text{THz}}$ is the product of $\zeta_{\text{STE}}$, an increasing function of $t_{\text{Pt}}$, times a product of two decreasing functions \textit{vs.} $t_{\text{Pt}}$ related to the NIR and THz absorptions (see Eq.~\textcolor{blue}{(ES1)} in the Suppl.~Mat.~\textcolor{blue}{S2}). The analysis yields an optimum of $t_{\text{Pt}}$ in both \textit{spin-sink} and \textit{hard-wall} systems.
The values of $t_{\text{Pt}}$ which yields a maximum $E_{\text{THz}}$ for different $\lambda_{sf}^{\text{Pt}}=1-10$~nm are given in Table~\ref{tabGSS} in the two cases. An optimal value of $t_{\text{Pt}}^{\text{max}}\lesssim 2$~nm for the \textit{spin-sink} limit on Fig.~\ref{fig2}\textcolor{blue}{b} well corresponds to $\lambda_{sf}^{\text{Pt}}\lesssim 3$~nm as previously determined. Table~\ref{tabGSS} also presents the effective reduction $\eta=\left( t_\text{Pt}^{\text{ss}} - t_\text{Pt}^{\text{hw}} \right) / t_\text{Pt}^{\text{hw}}$ of the optimal Pt thickness for the two models resulting in a coupled effect of \textit{spin-sink} electronic enhancement and reduction of the optical absorptions thus allowing a significant emission increase.

\begin{table}[!h]
\begin{tabular}{|c|c|c|c|c|c|c|c|}
\cline{1-8}
    {$\lambda_{sf}^\text{Pt}$}~(nm)& 1 & 2 & 3 & 4 & 5 & 7 & 10 \\ \cline{1-8}
    \hline\hline
    {$t_\text{Pt}^\text{ss}$}~(nm)& 1.1 & 1.7 & 2.1 & 2.4 & 2.6 & 3 & 3.4 \\ \cline{1-8}
    {$t_\text{Pt}^\text{hw}$}~(nm) & 2.2 & 3.3 & 4.1 & 4.6 & 5.3 & 6.2 & 7 \\ \cline{1-8}
    {$\mathcal{\eta}$~(\%)} & -50 & -48 & -48 & -48 & -51 & -52 & -51 \\ \cline{1-8}
\end{tabular}
\caption{Optimum values of the Pt thicknesses ($t_\text{Pt}$~in~nm) expressed \textit{vs.}~the spin diffusion length ($\lambda_{sf}^\text{Pt}$~in~nm) in case of the \textit{spin-sink} model and \textit{hard-wall} limit alongside the effective thickness reduction $\eta$ between the two models.}
\label{tabGSS}
\end{table}

\vspace{0.1in}

\begin{figure*}[!ht]
\begin{center}
\includegraphics[width=1\textwidth]{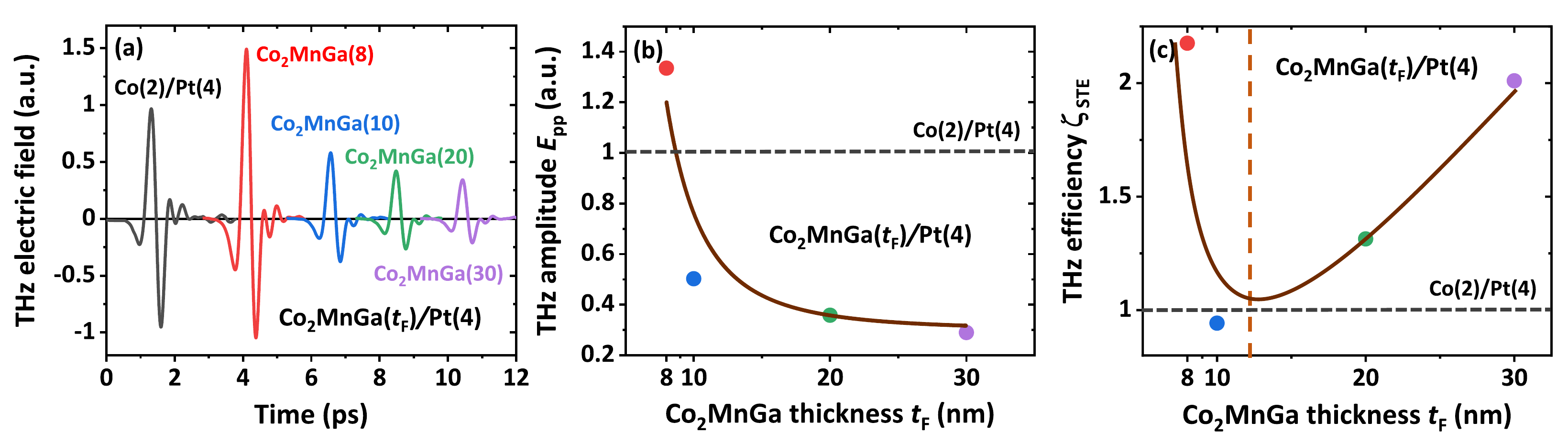}
\caption{\textbf{Co$_2$MnGa/Pt THz emission}. \textcolor{blue}{(a)} THz emission of Co$_2$MnGa($t_\text{F}$)/Pt as a function of the Co$_2$MnGa layer thickness $t_\text{F}$ (resp.~8,~10,~20,~30~nm) represented in the time-domain THz pulse and \textcolor{blue}{(b)} as peak-peak amplitude. Time traces are shifted in time for clarity. \textcolor{blue}{(c)} Renormalized THz emission taking into account the THz and NIR absorption in the heterostructures thus adressing $\zeta_\text{STE}$ electronic efficiency. The solid lines are a guide for the eyes.}
\label{fig3}
\end{center}
\end{figure*}

\vspace{0.1in}

Finally, we now turn to the role of the spin injector in the THz emission efficiency and particularly on the one of STEs based on magnetic Weyl half-metal Heusler alloys exceeding the one of the best CoFeB 3\textit{d} transition metal candidate. The integration of such members of the Heusler family, presently Co$_2$MnGa, in active STEs~\cite{schneider2021,Gupta2021,bierhance2021} has several potential advantages and few drawbacks. First, such magnetic Weyl materials are known to \textit{i}) behave as half-metals possessing a strong spin polarization connected to a majoritary spin-down reservoir~\cite{vasaprasad2010,guillemard2019,guillemardAPL} and \textit{ii}) possess a high transparency with Pt comparable to Co/Pt in connection to a large experimental spin-mixing conductance with Pt from spin-pumping~\cite{Chudo2011,Sasaki2020}. Moreover, Co$_2$MnGa has recently been demonstrated to carry a negative spin Hall angle $\simeq -20$\% (the sign is compared to Pt) owing to their spin-orbit and subsequent Berry phase~\cite{Leiva2021}. One may also expect strong THz emission of a single Co$_2$MnGa film (self-emission)~\cite{Wu2019} with a reduced conductivity ($\simeq$130~$\mu \Omega$.cm~\cite{Markou2019}) reducing the THz absorption based on the Drude model~\cite{Nadvornik2021}. All these arguments account for an ideal THz emission recipe. The samples investigated here were MgO//Co$_2$MnGa($t_\text{F}$)/Pt(4) structures (see Suppl.~Mat.~\textcolor{blue}{S1}).

We present in Figs.~\ref{fig3}\textcolor{blue}{a} and~\ref{fig3}\textcolor{blue}{b} the Co$_2$MnGa($t_\text{F}$)/Pt thickness dependence (resp.~8,~10,~20,~30 nm) of the THz emission. Notably, at $t_\text{F}$=8 nm, the generated THz pulse is still 35\% larger than the Co(2)/Pt(4) even if the spin-injector is four times thicker. This shows that Co$_2$MnGa is a more efficient spin-injector compared to Co and emphases the strong spin polarization of this Heusler compound. Moreover, the THz pulse generated from a single Co$_2$MnGa(43) layer (Suppl.~Mat.~\textcolor{blue}{S4}) is negligible compared to Co$_2$MnGa(43)/Pt bilayers. It may indicate the expected self-SCC contribution although its relative small amplitude. Figs.~\ref{fig3}\textcolor{blue}{a} and~\ref{fig3}\textcolor{blue}{b} show that an increase of the Co$_2$MnGa thickness reduces the emitted THz amplitude in line with optical absorption. To take into account these effects, we proceeded as previously to the THz and NIR renormalization procedure (see Suppl.~Mat.~\textcolor{blue}{S2}) in order to extract the STE electronic efficiency $\zeta_\text{STE}$. The tabulated results are presented in Fig.~\ref{fig3}\textcolor{blue}{c}.

This exhibits two large conversion efficiencies in two different $t_F$ thickness regions reaching twice that of Co/Pt at equivalent Pt thickness. For small $t_\text{F}$, the spin current arises from the ultrashort NIR pump and is converted to charge by ISHE in Pt demonstrating thus the high quality of the epitaxial Co$_2$MnGa thin films down to 8 nm. In this region, $\zeta_\text{STE}$ is enhanced from the increase of $r_\text{F}=\rho_{xx} l_{sf}^2/t_\text{F}\propto 1/t_\text{F}$~\cite{jaffres2010} in the expression of $\zeta_\text{STE}$ (Eq.~\eqref{zeta}). It corresponds to a spin-signal increase in a confined volume~\cite{lackzowski2012} and, in the present case, to a minimal spin relaxation in the spin-injector (\textit{resp.}~maximal spin relaxation in Pt). It leads to a full relaxation of excited electrons in Pt whatever the source term (either ultrafast pump absorption or thermal build-up). The second region of interest maps the increase of $\zeta_{\text{STE}}$ \textit{vs.} the Co$_2$MnGa thickness $t_\text{F}$. The presence of a sizeable SCC suggests a strong contribution from the Co$_2$MnGa bulk which may be a result of two effects~\cite{jimenezcavero2021}. Firstly, \textit{a}) a sizeable perpendicular thermal gradient generated by the NIR pump can lead to a transient spin-current generation \textit{via} anomalous Nernst effect as recently demonstrated by Reichlova~\textit{et.~al.}~\cite{reichlova2018}. The generated spin-current may flow from the hot interface with Pt towards the cold interface with MgO. Secondly, \textit{b}) the expected negative sign of the self-SCC of Co$_2$MnGa would lead to the generation of a THz wave in-phase with the one generated by ISHE in Pt, thus increasing $\zeta_{\text{STE}}$ at large $t_\text{F}$ (Fig.~\ref{fig3}\textcolor{blue}{c}).

 \vspace{0.1in}

To conclude, we have investigated a range of STEs based on FM/HM junctions and the ISHE for THz pulse generation. Using THz emission TDS, we show the optimization of three approaches of material and interface engineering to optimize the THz emission properties of STEs. This included \textit{i}) multi-stacks permitting a larger number of emitting dipole planes, and showed a compromise with the NIR and THz absorption with increasing stacks, \textit{ii}) the application and modelling of Au:W \textit{spin-sinks} to avoid detrimental reflections of hot spins at metallic interfaces and \textit{iii}) enhancing the optically generated spin-polarization and hence THz emission with the use of the semi-metals (Co$_2$MnGa) possessing a high spin-polarization. An important aspect of this work included the determination of the \textit{intrinsic} emission efficiency $\zeta_\text{STE}$ for THz pulse emission. These insights could provide further perspectives to the optimization of STEs where multi-stacks, \textit{spin-sinks} and semi-metals are combined.

\vspace{0.2in}

\section*{Data Availability Statements}
The data that supports the findings of this study are available within the article (and its supplementary material).

\vspace{0.1in}

\section*{Supplementary material}

See Supplementary Material for the experimental details of the growth, THz emission spectroscopy setup discussion, detailled electron diffusion modelling and THz emission renormalization procedure.

\vspace{0.1in}

\section*{Acknowledgements}

The authors thank G.~Bierhance and T.~Kampfrath for valuable discussions about Co$_2$MnGa/Pt based THz emission. We acknowledge the Horizon 2020 FETPROAC Project No.~SKYTOP-824123 “Skyrmion—Topological Insulator and Weyl Semimetal Technology”. We acknowledge financial support from the Horizon 2020 Framework Programme of the European Commission under FET-Open grant agreement No.~863155 (s-Nebula).

\bibliographystyle{ieeetr}
\bibliography{biblio}

\end{document}